\documentclass[osajnl,twocolumn,showpacs,superscriptaddress,10pt]{revtex4-1} 
\usepackage{amsmath,amssymb,graphicx}
\begin{document}

\title{Optomechanical steady-state entanglement induced by electrical interaction}

\author{ChunNian Ren}\email{scienceren@yeah.net}
\affiliation{College of Information Science and Engineering, Ocean University of
China, Qingdao 266100, China}

\author{JianQi Zhang}
\affiliation{State Key Laboratory of Magnetic Resonance and Atomic and Molecular
Physics,Wuhan Institute of Physics and Mathematics, Chinese Academy of
Sciences, and Wuhan National Laboratory for Optoelectronics, Wuhan 430071, China}
\author{Libo Chen}
\affiliation{School of Science, Qingdao Technological University, Qingdao 266033, China}
\author{YongJian Gu}\email{Corresponding author: yjgu@ouc.edu.cn}
\affiliation{College of Information Science and Engineering, Ocean University of
China, Qingdao 266100, China}

\begin{abstract}We propose a scheme for generating remote continuous steady-state entanglement of output light leaked from optomechanical system, in which two mechanical oscillators are coupled through long-range Coulomb interaction. we show that the entanglement of output light is affected by the detuning and the strength of the Coulomb interaction. We also demonstrate that two movable mirrors and two light beams can be entangled in the steady state. We suggest an experimental readout scheme to fully verify the characteristic of entangled state.
\end{abstract}


\maketitle 

\section{Introduction}

The research of micro-mechanical resonators (MRs) has attracted considerable
interests in both quantum mechanics and nano-technology in the past decade,
because of the fact that MRs is an ideal candiate to search quantum properties
on mescoscopic objects. These quantum properties not only provide insights
into the fundamental physical principle in quantum regime
\cite{MPootPhyRep2012}, but also give potential applications of MRs, such as,
opto-mechancial metrology \cite{FMarquardtPhysics2009}, quantum information
processing \cite{KStannigelPRL2012,prl-110-120503}, biological sensing
\cite{Nat.nanotech-3-501}, and gravitational wave detection \cite{omm}.

However, only few quantum properties on MRs can be achieved experimentally
directly, since the quantum properties on MRs are too weak to be observed, and
they are always covered by the thermal fluctuation. Moreover, limited by the
ground cooling condition \cite{measurementcooling}, only the MRs with high
frequencies can be directly cooled to its ground state with an average phonon
number $\left\langle n\right\rangle \ll1$ \cite{Nature-464-697}, and the MRs
with lower frequencies need to be cooled further to lower temperature.
Therefore, it is desirable to develop observing more quantum properties on
MRs. And the steady entanglement in MRs is one of this kind of quantum
properties. Quantum entanglement is one of the most important features of
quantum mechanics.

Until now, several entanglement schemes based on MRs have been proposed to
achieve continuous variable entanglement. These entanglment schemes range from
the cavity modes to MRs, including the entanglment for two MRs
\cite{Lingzhou2011,HartamannPRL2008,WenjieNie2012}, the entanglement between
one MR and a cavity mode in optomechancial cavity systems
\cite{DVitaliPRL2007}, entanglement of the output light with optomechancial
systems \cite{Lingzhou2011}, entanglement of the optical and microwave cavity
mode with a MR \cite{ShBarzanjehPRA2011}. However, this scheme is rarely
involved methods long-range entanglement\cite{TanhuatangPRA2013}.

The aim of our work is to generate the entanglement between the output light
fields leaking out of two sides of an optomechnical system with two charged
movible mirros in it. The Coulomb interaction between the two charged MRs in
such a system will set up the entanglement of the MRs and the modes in the
cavities, As the system reaches steady state, the output fields will be entangled.

Our scheme that the entanglement between the two movable oscillators and
between the two beams leaked from the two cavities created by coulomb
interaction is quitely different from the conventional optomechanical
system\cite{HartamannPRL2008,Lingzhou2011}which the entanglement between two
movable mechanical resonators is generated by the inner cavity modes or
induced by the external atoms. So our scheme belongs to a kind of new
structure for generation continuous entangled light. Contrast to the
conventional methodes, the coulomb interaction belong to long-range
interactions\cite{ULow1994PRL,DBohm1953PR}. Furthermore, when the leaking
beams are in entangle state, the two mirrors are cooled at the same time, so
the influence of external noise is small, The entanglement can be keep a
longer time coherence\cite{YingDanWangPRL2013}.

\section{Model and Hamiltonian}

\begin{figure}[htbp]
\centerline{\includegraphics[width=.8\columnwidth]{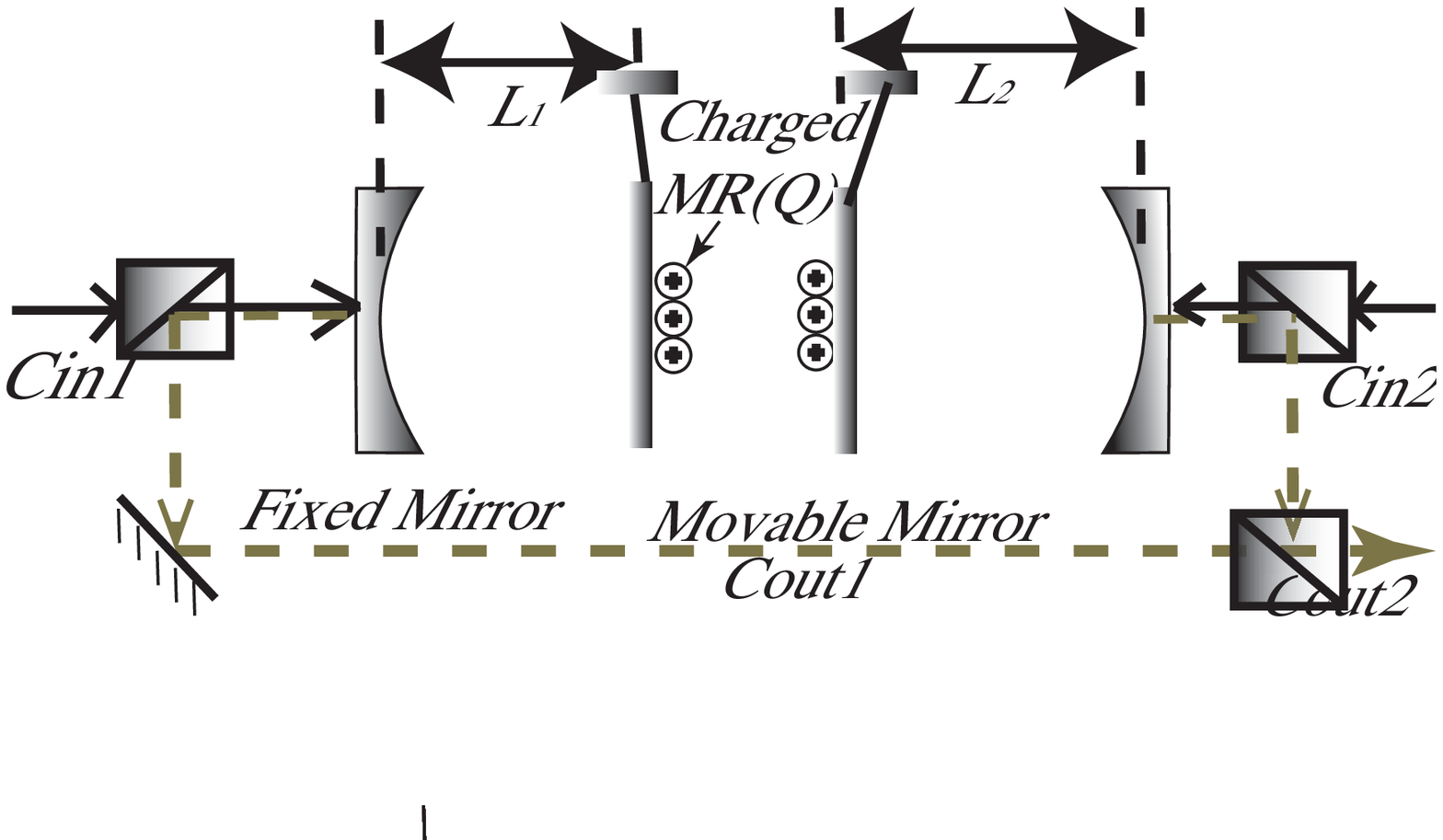}}
\caption{\label{fig:scheSys}Schematic description of the experimental system, including two
cavity. Each cavity with the length $L$ is driven by a classical light fields.
$r_{0}$ is the distance between the two movable mirrors in the absence of the
radiation pressure and the Coulomb force.}
\end{figure}

As it is sketched in Fig. \ref{fig:scheSys}, we consider the model is composed
of two spatially separated optomechanical cavities with a distance $r_{0}$.
Each opto-mechancial cavities consists of one fixed mirror and one charged MR.
When the distance between the two charged MRs is much large than the small
oscillations of the charged MRs $r_{0}\gg q_{m}$, the Coulomb interaction for
charged MRs can be written as $V=\lambda q_{1}q_{2}$
\cite{zhangjianqiPRA2012,pra-72-041405}, where $\lambda=\frac{2kQ_{1}Q_{2}%
}{r_{0}^{3}}$, $k$ is the electrostatic force constants, $Q_{m}$ is the net
charge for the MR $m$ ($=1,2$). After redefined the equilibrium position and
ignored the frequency shift caused by the Coulomb interaction, the motion of
the MRs can be given by $H_{MR}=\sum_{m=1}^{2}\frac{\hbar\omega_{m}}{2}\left(
p_{m}^{2}+q_{m}^{2}\right)  $, where $q_{m}$ and $p_{m}$ are the position and
momentum operators of MR $m$ with a frequency $\omega_{M}$. The energy for the
optomechancial cavities can be described as $H_{c}=\sum_{m=1}^{2}\hbar
\omega_{c,m}c_{m}^{\dag}c_{m}$ with $\omega_{c,m}$ being the frequency for the
cavity mode $c_{m}$. After each optomechancial cavities is driven by its
corresponding laser fields in the frequency $\omega_{p,m}$ with an input power
$P_{m}$ and a strength $\varepsilon_{p,m}=$ $\sqrt{2P_{m}/\hbar\omega_{c,m}}$,
the Hamiltonian describes our model can be give as:%
\begin{equation}%
\begin{array}
[c]{ccl}%
H_{T} & = & H_{c}+H_{MR}+V+H_{R}+H_{d};\\
H_{R} & = & \sum_{m=1}^{2}(-1)^{k}\chi_{m}q_{m}c_{m}^{\dag}c_{m};\\
H_{d} & = & \sum_{m=1}^{2}i\hbar\left(  \varepsilon_{p,m}e^{-i\omega_{p,m}%
t}c^{\dag}-H.c.\right)  ,
\end{array}
\label{OriTotHa}%
\end{equation}
where $H_{R}$\textbf{\ }is for the radiation pressure couplings beween the MR
and its corresponding cavity mode, where $\chi_{m}=\hbar ¦Ø _{c,m}/L_{m}$
is the strength of the radiation pressure coupling with a cavity length
$L_{m}$. The last item is describe the optomechancial cavity driven by the
external laser fields.

For simplify, we suppose both the optomechancial cavities and driven laser
fields are identical, then we can get $\varepsilon_{p,m}=\varepsilon_{p}$,
$\omega_{p,m}=$ $\omega_{p}$, and $\omega_{m}=\omega_{M}$. In the frame
rotating with the driving frequency $\omega_{p}$, we can rewritte the
Hamitlonian (\ref{OriTotHa}) as
\begin{equation}%
\begin{array}
[c]{rl}%
H_{T}= & \sum_{m=1}^{2}\left[  \hbar\Delta_{0}c_{m}^{\dag}c_{m}+\frac
{\hbar\omega_{M}}{2}\left(  p_{m}^{2}+q_{m}^{2}\right)  \right] \\
+ & \sum_{m=1}^{2}\left[  (-1)^{m}\chi q_{m}c_{m}^{\dag}c_{m}\right]  +\lambda
q_{1}q_{2}\\
+ & \sum_{m=1}^{2}i\hbar\left(  \varepsilon_{p}c_{k}^{\dag}-H.C.\right)  ,
\end{array}
\label{Eq:TotHam}%
\end{equation}
with $\Delta_{0}=\omega_{c}-\omega_{p}$ being the detuning from the cavity to
the laser field. Fig.~\ref{fig:scheSys}.

\section{Quantum Langevin equations}

A proper analysis of the system must include photon losses in the cavity and
the Brownian noise acting on the mirror. Substituting The total Hamiltonian in
Eq.(\ref{Eq:TotHam}) into that differential equation the Heisenberg equations
of motion and adding the corresponding damping and noise terms, we obtain the
quantum Langevin equations as follows:%
\begin{equation}%
\begin{array}
[c]{cc}%
\dot{q}_{m} & =\omega_{m}p_{m}\\
\dot{p}_{1} & =-\omega_{m}q_{1}+\chi c_{1}^{\dag}c_{1}+\lambda q_{2}%
-\gamma_{m}p_{1}+\xi_{1}\\
\dot{p}_{2} & =-\omega_{m}q_{2}+\chi c_{2}^{\dag}c_{2}+\lambda q_{1}%
-\gamma_{m}p_{2}+\xi_{2}\\
\dot{c}_{1} & =-\left(  \kappa+i\Delta_{0}\right)  c_{1}+i\chi q_{1}%
c_{1}+\varepsilon_{p_{1}}e^{-i\omega_{p_{1}}t}+\sqrt{\kappa}c_{1in}\\
\dot{c}_{2} & =-\left(  \kappa+i\Delta_{0}\right)  c_{2}+i\chi q_{2}%
c_{2}+\varepsilon_{p_{2}}e^{-i\omega_{p_{2}}t}+\sqrt{\kappa}c_{2in}%
\end{array}
\label{Eq:LangevinEq}%
\end{equation}
The quantum Brownian noise $\xi_{1}$and$\xi_{2}$are from the coupling of \ the
movable mirrors to their own own environment. We suppose the correlation
function at temperature $T$%
\begin{equation}%
\begin{array}
[c]{cc}%
\left\langle \xi_{j}\left(  t\right)  \xi_{k}\left(  t^{\prime}\right)
\right\rangle = & \frac{\delta_{jk}\gamma_{m}}{\omega_{m}}\int\frac{d\omega
}{2\pi}e^{-i\omega\left(  t-t^{\prime}\right)  }\omega\left[  1+\coth\left(
\frac{\hbar\omega}{2\kappa_{B}T}\right)  \right] \\
& j,k=1,2
\end{array}
\label{Noisephonon}%
\end{equation}
the mirror Brownian noise $\xi_{1}$ and $\xi_{2}$ are not Markovian and
therefore cann't be described by delta correlated
function\cite{DVitaliPRL2007,VGiovannettiPRA2001}. But the non-Markvian
effects are achievable only in the case that the oscillators are working in a
large mechanical quality factor $Q_{m}=\omega_{m}/\gamma_{m}\gg1$. Hence,
$\xi_{1}$ and $\xi_{2}$ become delta correlated:%
\[
\left\langle \xi_{j}\left(  t\right)  \xi_{k}\left(  t^{\prime}\right)
+\xi_{j}\left(  t^{\prime}\right)  \xi_{k}\left(  t\right)  \right\rangle
/2\simeq\gamma_{m}\left(  2\bar{n}+1\right)  \delta\left(  t-t^{\prime
}\right)
\]
here $\bar{n}=\left(  \exp\left\{  \hbar\omega_{m}/k_{B}T\right\}  \right)  $
the two cavity modes decay at the same rate $\kappa_{1}=\kappa_{2}=\kappa$,
and $a_{1in}\left(  a_{2in}\right)  $ is the vacuum radiation input noise with
the correlation relations which are given by%
\begin{equation}%
\begin{array}
[c]{cc}%
\left\langle a_{jin}^{\dag}\left(  t\right)  a_{jin}\left(  t^{\prime}\right)
\right\rangle  & =N\delta\left(  t-t^{\prime}\right) \\
\left\langle a_{jin}\left(  t\right)  a_{jin}^{\dag}\left(  t^{\prime}\right)
\right\rangle  & =\left(  N+1\right)  \delta\left(  t-t^{\prime}\right)
\end{array}
\label{NoiseOfPhoton}%
\end{equation}
here we sopposed the environment is a thermal equilibrium state , the photon
number $N=\left[  \exp\left(  \hbar\omega_{c}/k_{B}T\right)  -1\right]  ^{-1}%
$, here $k_{B}$ is the Boltzmann constant and $T$ is the mirror temperature.
In order to obtain the steady state of the eqation\ref{Eq:LangevinEq}, each
Heisenberg operator can always be rewritten as a $c$-number steady-state
equations plus an additional fluctuation operator with zero-mean value,
$\alpha=a_{s}+\delta a$, $q=q_{s}+\delta q$, $p=p_{s}+\delta p$. When we
insert these expressions into the Eqs.\ref{Eq:LangevinEq}, a set of nonlinear
algebraic equations for the steady state values and a set of quantum Langevin
equations for the fluctuation operators\cite{FabrePRA1994,DVitaliPRL2007} can
be calculated analytically . The values for the set of steady state equations
are read:%

\begin{equation}%
\begin{array}
[c]{cc}%
p_{1s} & =p_{2s}=0\\
q_{1s} & =\frac{\chi\left\vert c_{1s}\right\vert ^{2}+\lambda\chi\left\vert
c_{2s}\right\vert ^{2}}{\omega_{m}-\lambda^{2}}\\
q_{2s} & =\frac{\chi\left\vert c_{2s}\right\vert ^{2}+\lambda\chi\left\vert
c_{1s}\right\vert ^{2}}{\omega_{m}-\lambda^{2}}\\
c_{1s} & =\frac{\varepsilon_{p_{1}}}{\left(  \kappa+i\Delta-i\chi
q_{1s}\right)  }\\
c_{2s} & =\frac{\varepsilon_{p_{2}}}{\left(  \kappa+i\Delta-i\chi
q_{2s}\right)  }%
\end{array}
\label{steadyVal}%
\end{equation}

From Eq.\ref{steadyVal}, we have a third-order nonlinear equations array for
$\left\vert c_{ms}\right\vert $and $q_{m,s}$. Unfortunately, for the exact
expression is too cumbersome and will not be reported here. When radiation
pressure coupling is strong, significant optomechanical entanglement is
achieved\cite{DVitaliPRL2007}. For high finesse cavities and enough driving
power, the system is characterized by a semiclassical steady state with the
cavity mode in a coherent state with amplitude $\alpha_{s,m}=\varepsilon
_{_{p}in,m}/\left(  \kappa_{m}+i\Delta_{m}\right)  $, $\left\vert
\varepsilon_{_{p}in,m}\right\vert =$ $\sqrt{2P_{m}/\hbar\omega_{0,m}}$,
$m=1,2$. Using the approach of
\cite{SuMeiHuangNJP2009,HYLengPRA2009,LingZhouPRA2011}, The fluctuations can
be calculated analytically by solving the Eqs. Where we define a new opretor
$X_{m}=\left(  c_{m}+c_{m}^{\dag}\right)  /\sqrt{2}$and $Y_{m}=\left(
c_{m}-c_{m}^{\dag}\right)  i/\sqrt{2}$, $m=1,2$ and corresponding Hermitian
input noise operators $X_{m,in}=\left(  c_{m,in}+c_{m,in}^{\dag}\right)
/\sqrt{2}$and $Y_{m,in}=\left(  c_{m,in}-ic_{m,in}^{\dag}\right)  /i\sqrt{2},$
Then, we chose two new sets for the Langevin equation, one is $f=\left\{
\delta q_{1,}\delta p_{1},\delta q_{2},\delta p_{2},\delta X_{1},\delta
Y_{1},\delta X_{2},\delta Y_{2}\right\}  $ as the input vector and another is
$b=\left\{  0,\xi_{1},0,\xi_{2},\sqrt{\kappa}X_{1,in},\sqrt{\kappa}%
Y_{1,in},\sqrt{\kappa}X_{2,in},\sqrt{\kappa}Y_{2,in}\right\}  $the linear
Langevin equations. The optomechanical system will reaches a unique
steady-state as it is stable. In the steady state, the input light power is
very larger than the optomechanical entanglement in the parameter regime when
$\left\vert c_{s}\right\vert \gg1.$In this case, we can neglect the nonlinear
terms $\delta q\delta c$.%

\begin{subequations}
\label{Eq:fluctuation}%
\begin{align}
&
\begin{array}
[c]{cc}%
\delta\dot{q}_{1} & =\ \omega_{m}\delta p_{1}%
\end{array}
\\
&
\begin{array}
[c]{cc}%
\delta\dot{p}_{1} & =-\omega_{m}\delta q_{1}+F_{1}\delta X_{1}+\lambda\delta
q_{2}-\gamma_{m}\delta p_{2}+\xi_{1}%
\end{array}
\\
&
\begin{array}
[c]{cc}%
\delta\dot{q}_{2} & =\omega_{m}\delta p_{2}%
\end{array}
\\
&
\begin{array}
[c]{cc}%
\delta\dot{p}_{2} & =-\omega_{m}\delta q_{2}+F_{2}\delta X_{2}+\lambda\delta
q_{1}-\gamma_{m}\delta p_{2}+\xi_{2}%
\end{array}
\\
&
\begin{array}
[c]{cc}%
\delta\dot{X}_{1} & =G_{1}\delta Y_{1}+\sqrt{\kappa}X_{1,in}%
\end{array}
\\
&
\begin{array}
[c]{cc}%
\delta\dot{Y}_{1} & =G_{1}\delta X_{1}+F_{1}\delta q_{1}+\sqrt{\kappa}Y_{1,in}%
\end{array}
\\
&
\begin{array}
[c]{cc}%
\delta\dot{X}_{2} & =G_{2}\delta Y_{2}+\sqrt{\kappa}X_{2,in}%
\end{array}
\\
&
\begin{array}
[c]{cc}%
\delta\dot{Y}_{2} & =G_{2}\delta X_{2}+F_{2}\delta q_{2}+\sqrt{\kappa}Y_{2,in}%
\end{array}
\end{align}

Here we denote $\Delta=\kappa+i\Delta_{0}$, $G_{m}=\left(  \Delta-\chi
q_{s,m}\right)  $, $F_{m}=\sqrt{2}\chi\left\vert c_{s,m}\right\vert ^{2}$,
$m=1,2$.

As the quantum noise and $c_{m,in}$, $m=1,2$ are zero-mean quantum Gaussian
noise and the dynamics is linearized, the quantum steady state for the
fluctuations can be rewriten as the follow express:%

\end{subequations}
\begin{equation}
\dot{f}=Af+b \label{Eq:MatrixLanFluct}%
\end{equation}

Where $A$ is drift matrix\cite{Lingzhou2011} describing full character of the
quantum steady state for the fluctuations.%

\begin{equation}
A=\left(
\begin{array}
[c]{cccccccc}%
0 & -\omega_{m} & 0 & 0 & 0 & 0 & 0 & 0\\
-m\omega_{m}^{2} & 0 & -\lambda & -\gamma_{m} & -F_{1} & 0 & 0 & 0\\
0 & 0 & 0 & -\omega_{2} & 0 & 0 & 0 & 0\\
-\lambda & 0 & -m\omega_{2}^{2} & -\gamma_{m} & 0 & 0 & -F_{2} & 0\\
0 & 0 & 0 & 0 & 0 & G_{1} & 0 & 0\\
-F_{1} & 0 & 0 & 0 & G_{1} & 0 & 0 & 0\\
0 & 0 & 0 & 0 & 0 & 0 & 0 & G_{2}\\
0 & 0 & -F_{2} & 0 & 0 & 0 & G_{2} & 0
\end{array}
\right)  \label{Eq:CorrMatrix}%
\end{equation}

When all of the eigenvalues of matrix $A$ have a negative real parts, the
system is running in a stable and reaches states. Under the help of
Routh-Hurwitz criterion\cite{EXDeJesusPRA1987}, we can get the following three
nontrivial conditions on the system parameters:%

\begin{align}
\lambda^{2}-m_{1}m_{2}\omega_{1}^{2}\omega_{2}^{2}  &  >0\nonumber\\
G1\ast\left(  m_{2}F_{1}^{2}\omega_{2}^{2}-G_{1}\lambda^{2}+G_{1}m_{1}%
m_{2}\omega_{1}^{2}\omega_{2}^{2}\right)   &  >0\label{Eq:RouthHurwitz}\\
G_{1}G_{2}^{2}\left(  G_{1}\lambda^{2}-m_{2}F_{1}^{2}\omega_{2}^{2}-G_{1}%
m_{1}m_{2}\omega_{1}^{2}\omega_{2}^{2}\right)   &  >0\nonumber
\end{align}

The formal solution of Eq.\ref{Eq:MatrixLanFluct} is%
\begin{equation}
f\left(  t\right)  =M\left(  t\right)  f\left(  0\right)  +\int_{0}%
^{t}M\left(  s\right)  b\left(  t-s\right)  \label{Eq:AnalyticalSolution}%
\end{equation}

where$M\left(  t\right)  =\exp\left(  At\right)  $.

\section{Entanglement of the output field}

In order to analyze the nature of linear quantum correlations among the two
MRs and among the two beams output field, the steady state of the correlation
matrix of quantum fluctuations in this multipartite system can be considering.
The noises from the phonon bath and photon bath are both zero mean quantum
Gaussian noise, so the steady state of the system is a zero-mean multipartite
Gaussian state.
\subsection{Entanglement of the two mechanical oscillators
interacted by Coulomb force}

We use the defination $V_{ij}\left(  \infty\right)  =\frac{1}{2}\left[
\left\langle f_{i}\left(  \infty\right)  f_{j}\left(  \infty\right)
+f_{j}\left(  \infty\right)  f_{i}\left(  \infty\right)  \right\rangle
\right]  $ which is the element of the covariance matrix. The information of
entanglement of the two mirrors or two beams leaked from two sides of the
cavities can be obtain with the help of the covariance matrix. The literature
\cite{RSimonPRL2000, DuanPRL2000} had proposed two criteria of the continuous
variable entanglement. Here we used the Duan's criterion proposed in
\cite{DuanPRL2000} and developed by , a state is entangled if the summation of
the fluctuations in the two EPR-like operators $X$ and $Y$ satisfy the
following inequality: $\left(  \Delta X\right)  ^{2}+\left(  \Delta Y\right)
^{2}<2$. here $X_{m}=Q_{1}+Q_{2}$, $Y_{m}=P_{1}-P_{2}$. We focus on the
entanglement of the four possible bipartite subsystems of the four-body system
that can be formed by traceless the others degree of
freedom\cite{ShBarzanjehPRA2011}, such that we can obtain a reduced $4\times4$
CM $\tilde{V}$ from $V$.
\begin{equation}
V_{ij}=\sum_{k,l}\int_{0}^{\infty}ds\int_{0}^{\infty}ds^{\prime}M_{ik}\left(
s\right)  M_{jl}\left(  s^{\prime}\right)  \Phi_{kl}\left(  s-s^{\prime
}\right)  \label{Vij}%
\end{equation}
Here $\Phi_{kl}\left(  s-s^{\prime}\right)  =\left(  \left\langle b_{k}\left(
s\right)  b_{l}\left(  s^{\prime}\right)  +b_{l}\left(  s^{\prime}\right)
b_{k}\left(  s\right)  \right\rangle \right)  /2=D_{kl}\delta\left(
s-s^{\prime}\right)  $ is the matrix of the stationary noise correlation
functions. Here $D_{kl}=Diag([0,\gamma_{m}\left(  2\bar{n}_{m}+1\right)
,0,\gamma_{m}\left(  2\bar{n}_{m}+1\right)  ,$

$\kappa\left(  2\bar{n}_{c}+1\right)  ,\kappa\left(  2\bar{n}_{c}+1\right)
,\kappa\left(  2\bar{n}_{c}+1\right)  ,\kappa\left(  2\bar{n}_{c}+1\right)
])$ is a diagonal matrix and $\bar{n}_{m}=0,\bar{n}_{c}=0$. If we neglect the
frequency dependence The the frequency domain treatment is same to the time
domain and the Correlation matrices have the same form. Under the stability
conditions, The following equations for the steady-state can be obtained:
\begin{align}
M\left(  \infty\right)   &  =0\label{infM}\\
AV+VA^{T}  &  =-D \label{Eq:CM}%
\end{align}
Eq.\ref{Eq:CM} is named Lyapunov equation which is equivalent to the
Eq.\ref{Vij} for the steady-state. The linear equation for $V$ can be straight
forwardly solved used Eq.\ref{Eq:CM}; but the exact expression is too complex
to reported in the article. We used the logarithmic negativity $E_{N}$ as a
measure of entanglement\cite{FabrePRA1994, GVidalPRA2002, DBuonoJOSAB2010}.%
\begin{equation}
E_{N}=\max\left[  0,-\ln2\eta^{-}\right]  \label{LogNega}%
\end{equation}
here $\eta^{-}=\sqrt{2}\left\{  \sum\left(  V\right)  -\left[  \sum\left(
V\right)  ^{2}-4\det V\right]  ^{1/2}\right\}  ^{1/2}$, and $\sum\left(
\tilde{V}\right)  =\det B+\det C-2\det E$. $B,C$ and $E$ is the a $2\times2$
block form of the $V$:%
\[
V\equiv\left(
\begin{array}
[c]{cc}%
B & E\\
E^{T} & C
\end{array}
\right)
\]
The results are shown in Fig.2, where we study the entanglements of the two
mechanical oscillators at the steady state of the system versus the detuning
and for different values of the $\kappa,\gamma$ . Here we find a parameter
region close to that of recently performed optomechanical
experiments\cite{HMiaoNJP2012}, and for simplicity, choose all the parameters
of the two mirrors, two lasers and the two cavities to be the same.
$\omega/2\pi=10MHz$, $\kappa_{1}=\kappa_{2}=0.8\omega_{b}$, $\gamma_{m}%
/2\pi=100Hz$, $T=300mK$ , $\Delta_{0}=\omega_{b}$Figure 1 demonstrate

\begin{figure}[htbp]
\centerline{\includegraphics[width=.8\columnwidth]{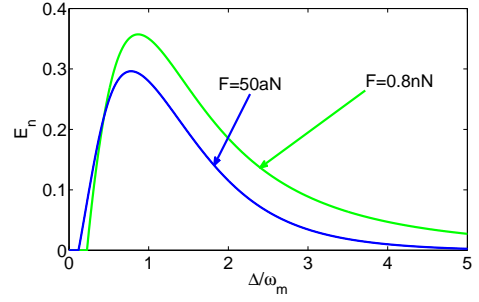}}
\caption{\label{fig:lognegativity}Plot of the logarithmic negativity $E_{N}$ as a function of the
normalized detuning $\Delta/\omega_{m}$ of the mechanical oscillator. Here the
optical cavity of length $L=25mm$, $P=50mW$, the mechanical oscillator has a
frequency $\omega_{m}/2\pi=10\ MHz$, a damping rate $\gamma_{m}/2\pi=100$
$Hz$.}
\end{figure}

$m=20ng$, $wavelength=\frac{2\pi c}{\omega_{c}}=1064nm$, $\omega_{c}%
/2\pi=2.8\times10^{14}Hz$,$C=27.5nF,U=1V,k_{q}=8.897Nm^{2}/C^{2}$ $F=0.88nN\symbol{126}%
10aN$Fig.~\ref{fig:lognegativity}.

\subsection{The continous entanglement between two beam
ouput light induced by optomechanical resonators}

In this section, we study the entanglement of the output field. The equation
in Eqs.\ref{Eq:LangevinEq} can be solved by the input-output relation:
\begin{equation}
\delta a_{jout}=\sqrt{2\kappa_{j}}\delta a_{j}-\delta a_{jin} \label{InOutR}%
\end{equation}

The Eq.\ref{Eq:MatrixLanFluct} can be written as $\delta\tilde{f}_{j}\left(
\omega\right)  =\left(  -i\omega-A\right)  ^{-1}b$ in the frequency domain by
Fourier transformation.

In order to study the nature of the output fields leaked from two sidebands of
our system, one can solve the analyzed solution from the two parts the
input-output relation for the two-mode field is the equations of
Eq(\ref{LangevinEq}) can be solved in the frequency domain by Fourier
transformation with the solution. Where $\delta\tilde{a}_{j}\left(
\omega\right)  $ is the Fourier transformation of $\delta a_{j}\left(
\omega\right)  $.

We only choose the part relevant to output fields from In the interaction
picture, $\omega$ represents the detuning from the cavity frequency. Used the
relation Eq.\ref{InOutR}, we obtain the following linear equation:%

\begin{equation}
\delta\tilde{f}_{out}(\omega)=c\left(  -i\omega-A\right)  ^{-1}b-e
\label{Eq:OutInFourier}%
\end{equation}

Where the matrices have the follow forms: $c=diag\left(  0,0,0,0,\sqrt
{2\kappa},\sqrt{2\kappa},\sqrt{2\kappa},\sqrt{2\kappa}\right)  $,
$e=[0,0,0,0,\delta X_{1in},\delta Y_{1in},\delta X_{2in},\delta Y_{2in}]^{T}$.
The output correlation matrix can be written as $V_{ij}^{out}\left(
\omega\right)  =\frac{1}{2}\left[  \left\langle f_{i}^{out}\left(
\omega\right)  f_{j}^{out}\left(  \omega^{\prime}\right)  +f_{j}^{out}\left(
\omega^{\prime}\right)  f_{i}^{out}\left(  \omega\right)  \right\rangle
\right]  $. $D_{kl}=Diag([\kappa\left(  2\bar{n}_{c}+1\right)  ,\kappa\left(
2\bar{n}_{c}+1\right)  ,\kappa\left(  2\bar{n}_{c}+1\right)  ,\kappa\left(
2\bar{n}_{c}+1\right)  ])$ The squeezing spectrum which is defined in a frame
of Fourer transformation can be calculated from the correlation matrix:%

\begin{equation}
\label{Cormatrix}%
\begin{array}
[c]{c}%
S_{out}\left(  \omega\right)  =\frac{1}{2}[\delta X_{f}\left(  \omega\right)
\delta X_{f}\left(  \omega^{\prime}\right)  +\delta X_{f}\left(
\omega^{\prime}\right)  \delta X_{f}\left(  \omega\right) \\
+\delta Y_{f}\left(  \omega\right)  \delta Y_{f}\left(  \omega^{\prime
}\right)  +\delta Y_{f}\left(  \omega^{\prime}\right)  \delta Y_{f}\left(
\omega\right)  ]
\end{array}
\end{equation}
In order to measure logarithmic negativity, one has to measure all independent
entries of the correlation matrix. We can used feasible experimental methods
have been realized in \cite{JLaurat}to experimental detection of the generated
entanglement of the output field. In our schematic, the measurement of the
field quadratures of the output field leaked from cavity can be
straightforwardly performed by homodyning the cavity output using a local
oscillator with an appropriate phase.

\section{Conclusion}

We propose a scheme to generate steady-state continuous entanglement of two
output beams which leaked from two sides of cavities induced by long-range
Coulomb interaction. We show that the entanglement of output light is affected
by the detuning and the strength of the Coulomb interaction. We also
demonstrate thate two movable mirrors and two light beams can be entangled in
the steady state. We suggest an experimental readout scheme to fully verify
the characteristic of entangled state. The results show that such
optomechanical entanglement can persist for higher environment temperatures using parameters based on the existing experiment.


\begin{thebibliography}{99}
\bibitem {MPootPhyRep2012}M. Poot and H. S. J. van der Zant, ``Mechanical systems in the quantum regime,'' Phys. Rep.
{\bf 511,} 273--335 (2012).

\bibitem {FMarquardtPhysics2009}F. Marquardt and S. M. Girvin, ``Trend:Optomechanics,'' Physics
{\bf 2,} 40--47 (2009).

\bibitem {KStannigelPRL2012}K. Stannigel, P. Komar, S. J. M. Habraken, S. D.
Bennett, M. D. Lukin, P. Zoller, and P. Rabl,``Optomechanical Quantum Information Processing with Photons and Phonons,'' \prl {\bf 109,} 013603--013607 (2012).

\bibitem {prl-110-120503}S. Rips, and J. Hartmann,``Quantum Information Processing with Nanomechanical Qubits,'' \prl {\bf 110,} 120503--120507 (2013).

\bibitem {Nat.nanotech-3-501}L. Tetard, A. Passian, K. T. Venmar, R. M. Lynch,
B. H. Voy, G. Shekhawat, V. P. Dravid, and T. Thundat,``Imaging nanoparticles in cells by nanomechanical holography,'' Nat. Nanotechnol.
{\bf 3,} 501--505 (2008).

\bibitem {omm}V. Braginsky and S. P. Vyatchanin,``Thermodynamical flutuations and photo-thermal shot noise in gravitational wave antennae,'' Phys. Lett. A {\bf 109,} 1-10 (1999).

\bibitem {measurementcooling}A. Schliesser ,O. Arcizet, R. Riviere, G. Anetsberger and
T. J. Kippenberg,``Resolved-sideband cooling and position measurement of a micromechanical oscillator close to the Heisenberg uncertainty limit,'' Nat Phys.{\bf 5,} 509--514 (2009).

\bibitem {Nature-464-697}O'Connell, A. D. Hofheinz, M.Ansmann, M.Bialczak,
Radoslaw C.Lenander, M.Lucero, Erik Neeley, M. Sank, D. Wang, H. Weides, M.
Wenner, J. Martinis, M. John,``Quantum ground state and single-phonon control of a mechanical resonatorNature ,''\nat  {\bf 464,}, 697--703 (2010).

\bibitem {Lingzhou2011}L. Zhou, Y. Han, J. T Jing, and W. P. Zhang, ``Entanglement of nanomechanical oscillators and two-mode fields induced by atomic coherence ,'' \pra  {\bf 83,}052117--052121 (2011).

\bibitem {HartamannPRL2008}M. J. Hartmann and M. B. Plenio,``Steady State Entanglement in the Mechanical Vibrations of Two Dielectric Membranes,''\prl
{\bf 101,} 200503-200506 (2008).

\bibitem {ULow1994PRL}U. L\H{o}w, V. J. Emery, and K. Fabricius,``Study of an Ising model with competing long- and short-range interactions,''\prl {\bf 72,} 1918--1921 (1994).

\bibitem {DBohm1953PR}D. Bohm and D. Pines,``Study of an Ising model with competing long- and short-range interactions,'' Phys. Rev. \textbf{92}, 609--625 (1953).

\bibitem {YingDanWangPRL2013}Y. D. Wang and A. A. Clerk,``Reservoir-Engineered Entanglement in Optomechanical Systems,''
 \pra {\bf 110,}
253601--253605 (2013).

\bibitem {WenjieNie2012}W. J. Nie, Y. H Lan, Y. Li, and S. Y Zhu,``Dynamics of a levitated nanosphere by optomechanical coupling and Casimir interaction,'' \pra {\bf 86,} 063809--063817 (2012).

\bibitem {ShBarzanjehPRA2011}Sh. Barzanjeh, D. Vitali, P. Tombesi, and G. J.
Milburn,`` Entangling optical and microwave cavity modes by means of a nanomechanical resonator,'' Phys. Rev. A. {\bf 84, } 063850--063860 (2011).

\bibitem {HYLengPRA2009}H. Y. Leng, J. F. Wang, Y. B. Yu, X. Q. Yu, P. Xu, Z.
D. Xie, J. S. Zhao, and S. N. Zhu,``Scheme to generate continuous-variable quadripartite entanglement by intracavity down-conversion cascaded with double sum-frequency generations,''  Phys. Rev. A. {\bf 79,} 032337--032346 (2009).

\bibitem {TanhuatangPRA2013}H. T. Tan, L. F. Buchmann, H. Seok, and G. X. Li,``Achieving steady-state entanglement of remote micromechanical oscillators by cascaded cavity coupling,''
Phys. Rev. A. {\bf 87,} 022318--022324 (2013).

\bibitem {KZhangPRL2012}K. Zhang, P. Meystre, and W. P. Zhang,``Role Reversal in a Bose-Condensed Optomechanical System
,'' \prl {\bf 108,} 240405--240409 (2012).

\bibitem {DVitaliPRL2007}D. Vitali, S. Gigan, A. Ferreira, H. R. B\"{o}hm, P.
Tombesi, A. Guerreiro, V. Vedral, A. Zeilinger, and M. Aspelmeyer,``Optomechanical Entanglement between a Movable Mirror and a Cavity Field ,''\prl {\bf 98,} 030405--030408 (2007).
\bibitem {TJKippenbergScience2008}T. J. Kippengberg and K. J. Vahala,``Cavity Optomechanics: Back-Action at the Mesoscale,'' Science
{\bf 321,} 1172--1176 (2008).

\bibitem {zhangjianqiPRA2012}J. Q. Zhang, Y. Li, M. Feng, Y. Xu,``Precision measurement of electrical charge with optomechanically induced transparency
,'' Phys. Rev. A
{\bf 86,}053806--053811 (2012).

\bibitem {pra-72-041405}W. K. Hensinger, D. W. Utami, H. S. Goan, K. Schwab,
C. Monroe, and G. J. Milburn,``Ion trap transducers for quantum electromechanical oscillators,'' Phys. Rev. A 72, 041405--041408 (2005).

\bibitem {VGiovannettiPRA2001}V. Giovannetti, and D. Vitali,``Phase-noise measurement in a cavity with a movable mirror undergoing quantum Brownian motion,''  Phys. Rev. A.
{\bf 63,} 023812--023819 (2001).

\bibitem {SumeiHuangNJP2009}S. M. Huang, and G. S. Agarwal,``Entangling nanomechanical oscillators in a ring cavity by feeding squeezed light
,'' New J. Phys.
{\bf 11,} 103044--103057 (2009).

\bibitem {EXDeJesusPRA1987}E. X. DeJesus, and C. Kaufman,``Routh-Hurwitz criterion in the examination of eigenvalues of a system of nonlinear ordinary differential equations
,'' Phys. Rev. A.
{\bf 35,} 5288--5290 (1987).

\bibitem {RSimonPRL2000}R. Simon,``Peres-Horodecki Separability Criterion for Continuous Variable Systems
,''Phys. Rev. Lett. {\bf 84,} 2726--2729 (2000).

\bibitem {DuanPRL2000}L. M. Duan, G. Giedke, J. I. Cirac, and P. Zoller,``Inseparability Criterion for Continuous Variable Systems,'' Phys.
Rev. Lett. {\bf 84,} 2722--2725 (2000).

\bibitem {GVidalPRA2002}G. Vidal and J. I. Cirac, ``Nonlocal Hamiltonian simulation assisted by local operations and classical communication
,'' Phys. Rev. A. {\bf 66,}
022315--022326 (2002).

\bibitem {GAdessoPRA2004}G. Adesso, A. Serafini, and F. llluminati,`` Extremal entanglement and mixedness in continuous variable systems,'' Phys. Rev. A. {\bf 70,} 022318--022335 (2004).

\bibitem {FabrePRA1994}C. Fabre, M. Pinard, S. Bourzeix, A. Heidmann, E.
Giacobino, and S. Reynaud,``Quantum-noise reduction using a cavity with a movable mirror
,'' Phys. Rev. A. {\bf 49,} 1337--1343 (1994).

\bibitem {DBuonoJOSAB2010}D. Buono, G. Nocerino, V. D'Auria, A. Porzio,
S.Olivares, and M. G. A. Paris,``Quantum characterization of bipartite Gaussian states
,'' J. Opt. Soc. Am. B {\bf 27,} 00A110 (2010).

\bibitem {HMiaoNJP2012}H. Miao, K. Srinivasan, and V. Aksyuk,`` A microelectromechanically controlled cavity optomechanical sensing system
,'' New J. Phys.
{\bf 14,} 075015--075031 (2012).

\bibitem {JLaurat}J. Laurat,``Entanglement of two-mode Gaussian states: characterization and experimental production and manipulation,''  J. Opt. B {\bf 7,} s577--s588 (2005).

\bibitem {Genes2008}C. Genes, A. Mari, P. Tombesi, and D. Vitali,``Robust entanglement of a micromechanical resonator with output optical fields,''Phys. Rev.
A. {\bf 8,} 032316--032329 (2008).

\end{thebibliography}
\end{document}